\documentclass[twocolumn,showpacs,preprintnumbers,amsmath,amssymb]{revtex4}
\usepackage{graphicx}
\usepackage{epstopdf}
\usepackage{amssymb}
\usepackage{dcolumn}
\usepackage{bm}

\begin{document}


\title{A multiobjective optimization approach to statistical mechanics}

\author{Lu\'is F. Seoane$^{1,2}$ and Ricard Sol\'e$^{1,2,3}$ }
\affiliation{  
    \\ $^1$ ICREA-Complex  Systems   Lab,  Universitat Pompeu Fabra  -  PRBB,
    Dr. Aiguader 88, 08003  Barcelona, Spain
    \\ $^2$ Institut de Biologia Evolutiva, UPF-CSIC, Passg Barceloneta, 08003 Barcelona
    \\$^3$ Santa Fe Institute, 1399 Hyde Park Road, New Mexico 87501, USA
  }

\begin{abstract}

  Optimization problems have been the subject of statistical physics approximations. A specially
relevant and general scenario is provided by optimization methods considering tradeoffs between cost
and efficiency, where optimal solutions involve a compromise between both. The theory of Pareto (or
multi objective) optimization provides a general framework to explore these problems and find the
space of possible solutions compatible with the underlying tradeoffs, known as the {\em Pareto
front}. Conflicts between constraints can lead to complex landscapes of Pareto optimal solutions
with interesting implications in economy, engineering, or evolutionary biology. Despite their
disparate nature, here we show how the structure of the Pareto front uncovers profound universal
features that can be understood in the context of thermodynamics. In particular, our study reveals
that different fronts are connected to different classes of phase transitions, which we can define
robustly, along with critical points and thermodynamic potentials. These equivalences are
illustrated with classic thermodynamic examples.

\end{abstract}

\maketitle

  \section{Introduction}
    \label{sec:1}
    
    Most complex systems result either from evolutionary or design processes where multiple
constraints must be simultaneously satisfied \cite{Schuster2012}. This includes living as well as
technological and economic systems \cite{Murray1926, WestEnquist1997, WestEnquist1999,
GafiychukLubashevsky2001, GastnerNewman2006, CuntzSegev2007, PerezEscuderoPolavieja2007,
CarvalhoArrowsmith2009, BassettBullmore2010, HasenstaubSejnowski2010, CuntzHausser2010,
Barthelemy2011}, in which very often the costs of implementing a task are confronted to its
efficiency. More complicated scenarios might involve other conflicting traits as well. Optimization
stands as a unifying principle that brings together questions from distant fields. The parallelisms
are further highlighted by the many engineering problems that are addressed through computer-based
Darwinian processes, or by the suggestion that biology {\em is} engineering fueled by natural
selection \cite{Dennett1995}.

    The simultaneous optimization of several traits is usually known as {\em Pareto} or {\em Multi
Objective Optimization} (MOO, \cite{Schuster2012, Coello2006}). Given a set $X$ composed of
candidate designs or solutions $x \in X$, the challenge is to find those elements $$x_\pi \in \Pi
\subset X$$ that simultaneously optimize a series of traits $$T_f = \{t_1, \dots, t_K\}$$ A simple
example is provided by the problem of how to use a given amount of money in improving a old car. We
might want to change it externally (painting and polishing the chassis) or instead invest and
improve the engine. Since money places a limit to what can actually be done, the user needs to
decide what type of compromise is to be achieved. In this way, the solution of an MOO ($\Pi \subset
X$, the Pareto front) comprises the most optimal tradeoff between the different targets.

    Previous works have been dedicated to study Pareto optimal sets \cite{FonsecaFleming1995,
Dittes1996, Zitzler1999, KonakSmith2006} particularly in engineering and economy \cite{Arrow1963,
SteuerNa2003, Coello2006} . Recent work includes diverse biological systems
\cite{CutelloNicosia2006, DruckmannSegev2007, Kennedy2010, GrimePapaspyrou2012, NoorMilo2012,
ShovalAlon2012, SchuetzSauer2012}, network models \cite{GoniSporns2013, PriesterPeixoto2014},
regulatory circuits \cite{HigueraMora2012, SzekelyAlon2013, OteroMurasBanga2014}, or control theory
of complex dynamics \cite{WilsonMiquel2004, DenysiukTorres2014}. All are case-dependent, and no
search for universal principles is made. However, insights into universal properties emerge as soon
as we compare MOO with the physics of phase transitions. In many systems, ordered and disordered
phases are separated by a transition point. Order shall result from energy minimization processes
that favor neighboring particles adopting the same state. Disorder, instead, arises from noise that
interfere with local ordering \cite{Jaynes1957,Harte2011, MoraBialek2011}. These transitions have
been identified in physical, social and economic systems \cite{Ball2004, CastellanoLoreto2009}. They
fall under two main classes (first or second order transitions) and display universal properties. If
we can frame a MOO problem within the context of phase transitions, several powerful elements of
this theory allow us to derive analytic results. In this context, statistical physics has been used
in optimization problems involving a global fitness function \cite{Fontana1989, PrugelShapiro1994,
AsselmeyerRose1997, Bornholdt1998, MartinZecchina2001, Bialek2013} and their associated phase
transitions \cite{HoggWilliams1996, GentWalsh1996, ReimannEbeling2002, Banga2008,
NeherShraiman2011}. However, little attention has been paid to the physics of Pareto optimality and
in many cases multiobjective problems are reduced to a Single Objective Optimization (SOO) using
global {\em fitness} or {\em energy functions} such as:
      \begin{eqnarray} 
        \Omega (t_1, ..., t_K; \lambda_1, ..., \lambda_K) &=& \sum_k \lambda_k t_k.
        \label{eq:1.01}
      \end{eqnarray} 
This approach has been used within the analysis of complex networks \cite{MathiasGopal2001,
FerrerSole2003a, ColizzaRinaldo2004, GastnerNewman2006, Newman2010} and models of human language
\cite{FerrerSole2003b, PropopenkoPolani2010, DickmanAltmann2012, SoleSeoane2014}. However, different
$t_k$ might not be commensurable implying that the parameters $\{\lambda_j\}$ introduce arbitrary
biases. We will avoid this linear integration initially, which is why we turn to Pareto optimality.
Thanks to this we will uncover a deep connection with key thermodynamic objects \cite{Gibbs1873a,
Gibbs1873b, Maxwell1904}. That will reveal a series of universal features across Pareto optimal
systems. These features are tightly mapped to phase transitions (including thermodynamic ones) when
the linear approach is recovered.

  \section{Methods}
    \label{sec:2}

    \subsection{Multiobjective optimization in a nutshell}
      \label{sec:2.01}

\begin{figure*}[htbp]
          \begin{center}

            \includegraphics[width = 15 cm]{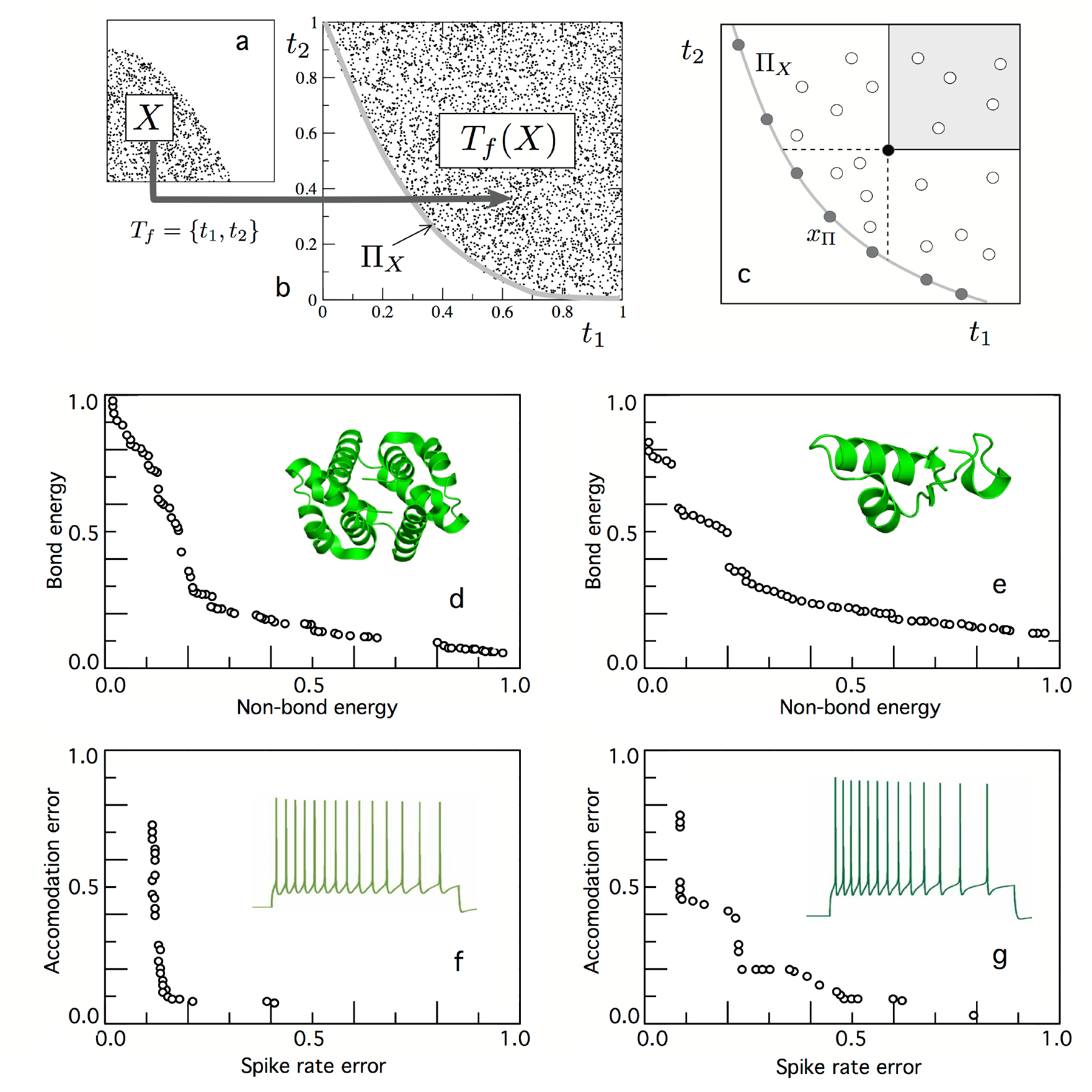}
            \vspace{0.4 cm}

            \caption{\textbf{Pareto optimality and Pareto front diversity} (a) A general picture of
the way the Pareto front $\Pi_X$ is constructed by mapping from the space $X$ to the target space
(b) where a new cloud of points $T_f(X)$. The boundary of this cloud defines $\Pi_X$ (gray line).
Within the previous cloud, different types of solutions can be identified. In (c) we indicate with
an arbitrary filled circle a non-optimal solution that dominates some others (within the fray
square) and is dominated by others, delimited between the two dashed lines. All open circles involve
non- Pareto optimal solutions. The Pareto front defines the limits of what is feasible. In (d-e) we
show two examples of Pareto fronts obtained from evolved protein structures (adapted from
\cite{CutelloNicosia2006}) and in (f-g) two additional examples of evolved firing neuronal patterns
to fit experimental data (adapted from \cite{DruckmannSegev2007}). For convenience, we have
normalised all axes. Notice the diverse structure of the fronts and the presence of changing
curvatures, gaps and kinks. }

           \label{fig:1}
          \end{center}
        \end{figure*}

      This section provides a summary of key aspects of MOO that can be found in the literature
\cite{Coello2006, FonsecaFleming1995, Dittes1996, Zitzler1999, KonakSmith2006}. Hereafter (and
without loss of generality) we assume minimization. Consider a set $X$ of objects upon which a
minimization will be enforced (figure \ref{fig:1}{\bf a}). We refer to all objects $x \in X$ as {\em
feasible} or {\em candidate solutions} or {\em designs}. Objects outside $X$ are {\em not feasible}.
Among all objects $x \in X$ we wish to find the subset $\Pi \subset X$ that minimizes a series of
given, real valued mathematical features (our target functions mentioned above):
        \begin{eqnarray}
          T_f(x) &\equiv& \{ t_k(x) ; k=1, ..., K \}, 
        \end{eqnarray} 
Our task is to find those objects that score lower in all $t_k \in T_f$ simultaneously.  $T_f$
establishes a mapping between $X$ and $\mathbb{R}^K$, that we refer to as {\em target space} (figure
\ref{fig:1}{\bf b}).

      We say that a candidate solution $x$ dominates another $y$ (and we denote it $x\prec y$) if
$t_k(x) \le t_k(y)$ for every $k=1,...,K$ and there is at least one $k'$ such that $t_{k'}(x) <
t_{k'}(y)$ (figure \ref{fig:1}{\bf c}). If a solution dominates another it is \emph{objectively
better} -- i.e. \emph{more optimal} -- considering {\em all} targets. A solution $x\in X$ will be
Pareto optimal if it does not exist any other feasible solution $y\in X$ such that $y\prec x$. The
Pareto front is the set of all Pareto optimal solutions ($x_\pi \in \Pi$) of an MOO. This object is
mapped into $\mathbb{R}^K$ by $T_f$ yielding a hypersurface $T_f(\Pi)$ of dimension $K-1$ or lower
(figure \ref{fig:1}{\bf b}). We shall refer to $T_f(\Pi)$ or $\Pi$ as Pareto optimal set or Pareto
front indistinctly.

      In figure \ref{fig:1}{\bf b-c} we plotted a very smooth front, but more complex ones show up
in real examples. Consider a Pareto optimality approach to protein structure prediction
\cite{CutelloNicosia2006} (figure \ref{fig:1}{\bf d-e}). Atoms in a protein are subjected to both
local forces (through their bounds with neighboring atoms) and mean-field forces such as the van der
Waals potential emerging from coarse grained, distant atoms. Both contributions should be minimal at
equilibrium. The Pareto efficient set of protein structures optimally trades between local and
global forces and produces an effective ensemble of protein conformations. In the same vein, we
could search for models of spiking neurons that simultaneously minimize their divergence with
respect to different aspects of real spike trains, as in \cite{DruckmannSegev2007} (figure
\ref{fig:1}{\bf f-g}). This again produces ensembles of models. In these and other cases (see
\cite{GoniSporns2013, PriesterPeixoto2014, DenysiukTorres2014}) we notice that the fronts display
complex shapes, suggesting inhomogeneous accessibility to different alternative optimal solutions.
As shown below, these changes in the Pareto front structure result from the presence of phase
transitions.

  \subsection{Simplest SOO defined upon an MOO -- MOO-SOO collapse}
    \label{sec:2.02}

    Given an MOO with $K$ target functions $T_f\equiv\{t_k, k=1,...,K\}$ we define the simplest SOO
problem by a linear combination of the targets through a set of external parameters
$\Lambda\equiv\{\lambda_k; k=1,...,K\}$. This produces a global {\em energy function}:
        \begin{eqnarray}
          \Omega(x, \Lambda) &=& \sum_k \lambda_k t_k(x). 
          \label{eq:2.01.01}
        \end{eqnarray}
This {\em energy} is an analogy, but since $\Omega(x, \Lambda)$ is minimized, global optima dwell at
the minimums of a potential landscape, which results in very intuitive visualizations of our optimal
systems.

    The minimization of $\Omega$ for a given $\Lambda$ with fixed values $\lambda_k \in \Lambda$
yields one SOO problem, thus equation \ref{eq:2.01.01} defines a parameterized family of SOOs. We
will study i) these SOOs, ii) the constraints that the Pareto front imposes to their solutions, and
iii) the relationships between different SOOs of the same family. The validity of the results holds
for any positive, real set $\Lambda$ but for convenience: i) We take $K=2$, which simplifies the
graphic representations and contains the most relevant situations possible. Some remarks are given
about $K>2$. ii) We require $\sum_k \lambda_k=1$ without loss of generality. For $K=2$ then
$\lambda_1=\lambda$, $\lambda_2=1-\lambda$, and $$\Omega = \lambda t_1 + (1-\lambda)t_2$$iii) We
impose $\lambda_k\ne0\>\>\forall k$. The case $\lambda_k=0$ is briefly commented.

\begin{figure}[htbp]
          \begin{center}
            \includegraphics[width = 8 cm]{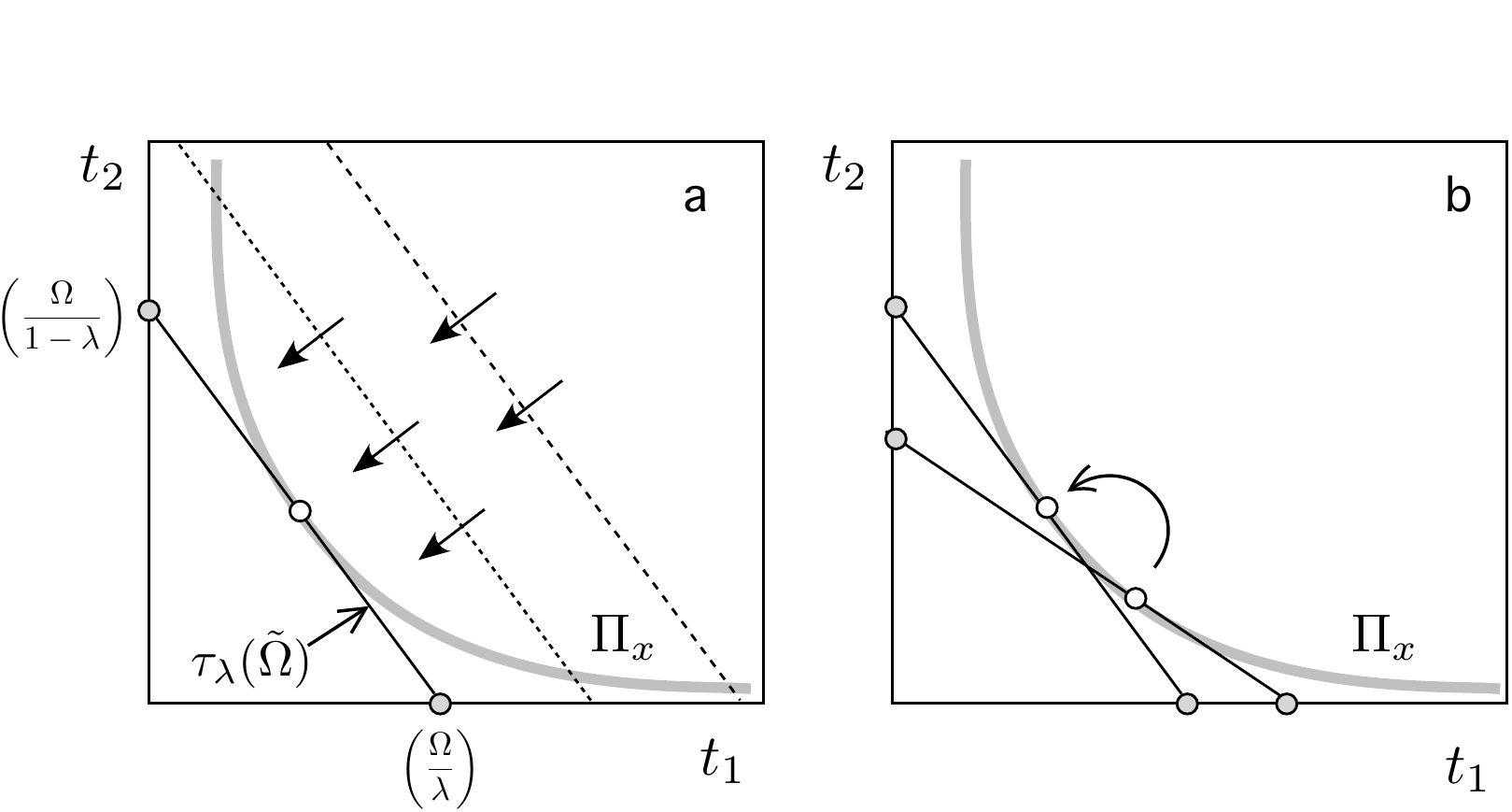}

            \caption{{\bf From single to multi-objective optimization}. (a) When a single-object
optimization assumption is made, a linear relation is defined between the target functions through
an energy function $\Omega(\lambda)=\lambda t_1 + (1-\lambda) t_2$ (see text). For a fixed $\lambda$
one sole SOO is posed whose solution lies where $\tau_\lambda(\Omega)$ (straight lines with slope $d
= -\lambda/(1-\lambda)$) matches the tangent of the front. (b) By changing $\lambda$, we visit other
solutions of the same SOO family. }

            \label{fig:2}
          \end{center}
        \end{figure}

    For given $\lambda_k$, one definite SOO problem is posed. Then, equation \ref{eq:2.01.01} with
fixed $\Omega$ defines {\em isoenergetic} surfaces noted $\tau_\Lambda(\Omega)$. Each
$\tau_\Lambda(\Omega)$ constitutes a $K-1$ dimensional hyperplane in the $K$-dimensional target
space. For $K=2$ (figure \ref{fig:2}{\bf a}) these surfaces are defined as:
      \begin{eqnarray}
        \tau_\lambda(\Omega) \equiv \left\{(t_1, t_2 ) \>\> \vert \>\> t_2 = {\Omega\over 1-\lambda} - {\lambda \over 1-\lambda}t_1 \right\}. 
        \label{eq:2.01.02}
      \end{eqnarray}
This $\tau_\lambda(\Omega)$ for $K=2$ means that, for a fixed $\lambda$, all solutions laying on the
same straight line defined by equation \ref{eq:2.01.02} have the same energy $\Omega$. Solutions
with lower or higher values of $\Omega$ for the same $\lambda$ lay also in straight lines parallel
to the original one. For general $K \ge 2$, the slope of $\tau_\Lambda (\Omega)$ along each possible
direction $\hat{t}_k$ in the target space only depends on $\Lambda$ so that different
$\tau_\Lambda(\Omega)$ for a given SOO problem are parallel to each other. In particular, for $K=2$
from equation \ref{eq:2.01.02}, we read the slope $${dt_2 \over dt_1} = -{\lambda \over 1-\lambda}$$

    The crossing of $\tau_\Lambda (\Omega)$ with each axis $\hat{t}_k$ is proportional to $\Omega$.
(figure \ref{fig:2}{\bf a}) With $\lambda_k$ given and constant, minimizing $\Omega$ means
finding $\tau_\Lambda(\tilde{\Omega})$ with $\tilde{\Omega}$ the lowest value possible such that
$\tau_\Lambda(\tilde{\Omega})$ still intersects the Pareto front. Graphically, this is equivalent to
{\em pushing} the isoenergetic surfaces against the Pareto front as much as possible (figure
\ref{fig:2}{\bf a}). Hyperplanes with lower $\Omega$ exist, but the Pareto front sets the
limit of feasibility: any solution with $\Omega<\tilde{\Omega}$ cannot be physically realized. The
SOO optimum always lays on the Pareto front. (Take $z \notin \Pi$, then $\exists x \in \Pi,
\>\>x\prec z$; thus at least for one $k'\in\{1,...,K\}$ we have $t_{k'}(x) < t_{k'}(z)$, implying
$\Omega(x, \Lambda) < \Omega(z, \Lambda)$ and $z$ cannot be SOO optimal.)

      The SOO optimum usually lays at the point $x_\Lambda \in \Pi$ where $\tau_\Lambda
(\tilde{\Omega})$ is tangent to the Pareto front (figure \ref{fig:2}{\bf a}). Exceptions to this
constitute the most interesting cases. The solution to different SOOs (defined by different values
of $\lambda$) are found in different points along the front (figure \ref{fig:2}{\bf b}). The
relationships within a family of SOO problems is thus partly encoded in the surface geometry.

  \subsection{Concavity/convexity and order parameters}
    \label{sec:2.03}

    The results that follow rely on a notion of concavity and convexity. The target surfaces
$\tau_\Lambda (\Omega)$ introduce a preferred direction along which minimization proceeds. This
provides a notion of {\em more} and {\em less} optimal ({\em lower} and {\em higher} energy) so that
concavities and convexities are consistently defined.

    Besides, the following results deal with phase transitions that are reflected in {\em order
parameters}. These represent some quality of our optimal designs that varies as a response to {\em
control parameters} -- these will be the biases $\{\lambda_j\}$. When phase transitions are present,
these responses happen in very characteristic ways.

    As for order parameters, we admit any physical, geometrical, topological, or any other features
that we can measure on the elements of $X$. We just require i) that they make different phases
distinguishable (they would be poor order parameters otherwise) and ii) that  non-trivial behaviors
arise out of the optimization dynamics exclusively -- if not, we might encounter order parameters
that become singular for some mathematical reason not relevant to our study. If the chosen
indicators obey these conditions, then the singularities that we call phase transitions arise {\em
for all} order parameters simultaneously.

    For an arbitrary order parameter $\theta$ the first condition implies that if $x,y\in X$ are
mapped into the same point ($T_f(x) = T_f(y)$), then $\theta(x)=\theta(y)$. The opposite ($\theta(x)
\ne \theta(y) \Rightarrow T_f(x) \ne T_f(y)$) is only required whenever $x\nprec y$ and $y\nprec x$.
This last condition guarantees that two points with different values of the order parameter are
never mapped into the same point of the Pareto front in $\mathbb{R}^K$. The second condition is
satisfied for $\theta$ such that $x,y\in X$ with $T_f(x) = T_f(y)+D\mathbb{R}^K$ implies $\theta(x)
= \theta(y)+D\theta$, $D\mathbb{R}^K$ and $D\theta$ standing for arbitrary differential
modifications. Then $\theta$ will not present non-analyticities other than those revealed by the
theory above. Following these conditions, the $t_k(x)$ themselves are valid order parameters.

  \section{Results}
    \label{sec:3}

    \subsection{Importance of the shape of the Pareto front}
      \label{sec:3.01}

      The most simple interplay between our MOO and the corresponding family of SOOs happens when
the Pareto front is convex and its tangent in the $\hat{t}_1 - \hat{t}_2$ plane is well defined in
its interior and its slope spans the interval $(-\infty,0)$ (figure \ref{fig:2}{\bf a}). Then,
the solution to the SOO posed by a given $\lambda$ is always found where the Pareto front has slope
$d=-\lambda/(1-\lambda)$ and $\tau_{\lambda} (\tilde{\Omega})$ matches the tangent of the front. A
differential increase $\lambda \rightarrow \lambda + D\lambda$ modifies the slope: $$d\rightarrow d
+ {2\lambda - 1 \over (1 - \lambda)^2}D\lambda$$ of the $\tau_\lambda(\tilde{\Omega})$. For $\lambda
\in (0,1)$, $d \in (0,-\infty)$. Each $\lambda$ poses an SOO with a different solution. Varying
$\lambda$, successive SOO solutions {\em roll smoothly} over the front (figure \ref{fig:2}{\bf
b}). This is similar to laying a rigid straight line ($\tau_\lambda(\tilde{\Omega})$, indeed)
against the front and reading the solution for different inclinations of the
$\tau_\lambda(\tilde{\Omega})$ at the contact point between that rigid line and the front. Any order
parameter $\theta$ renders a continuous, differentiable function of $\lambda$.

    \paragraph{Second order phase transitions}

\begin{figure*}[htbp]
          \begin{center}
            \includegraphics[width= 15 cm]{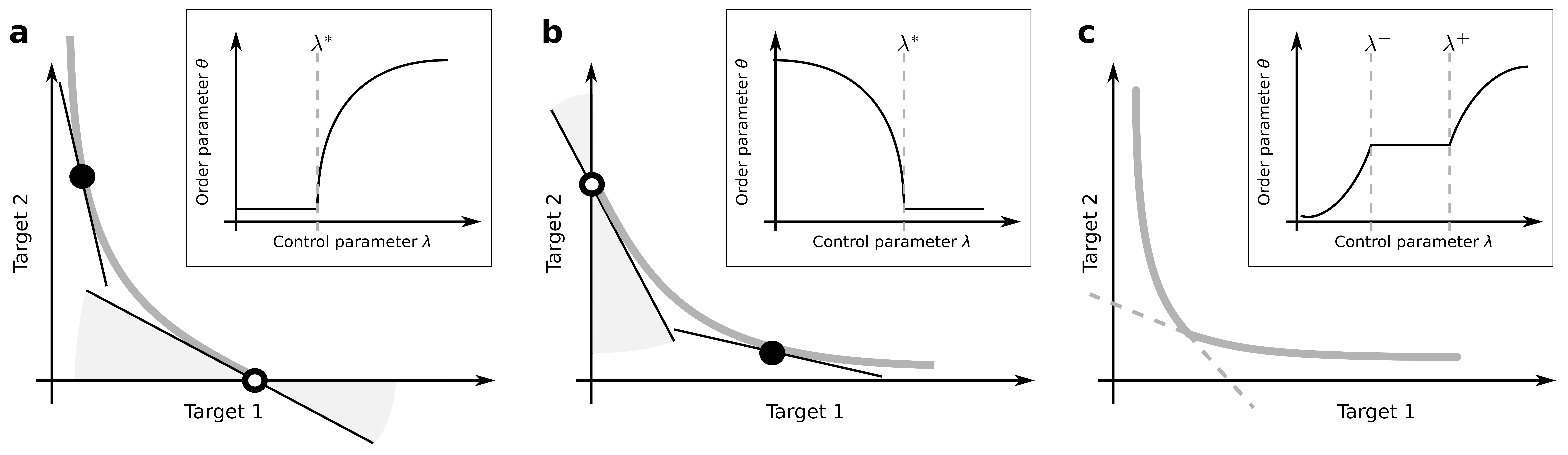}

            \caption{\textbf{Convex Pareto front with a tangent whose slope does not span the whole
range $(-\infty,0)$.} {\bf a} The slope of the Pareto front spans $d \in (-\infty,d^*)$. The front
{\em ends abruptly} at its bottom-right. There is a range ($\lambda < \lambda^*$, with $\lambda^* =
-d^*/(1-d^*)$ indicated by the gray fan) for which well defined SOO problems exist, whose solution
is persistently the same (open circle). For $\lambda>\lambda^*$ (filled circle) the front is sampled
gently as before. Any order parameter $\theta$ (inset) does not change if $\lambda < \lambda^*$
because the SOO optimum remains unchanged. Its derivative is not zero for $\lambda>\lambda^*$. This
causes an abrupt shift in ${d\theta\over d\lambda}$ at $\lambda^*$ while $\theta(\lambda)$ remains
continuous. {\bf b} The exact same situation happens if the pathology is found at the top- left of
the front. {\bf c} A sharp edge is associated with two discontinuities in the derivative of any
order parameter. }

            \label{fig:3}
          \end{center}
        \end{figure*}

      In figure \ref{fig:3}{\bf a, b}, and {\bf c} we represent convex Pareto fronts whose slopes
span $d\in(-\infty, d^*)$, $d\in(d^*, 0)$, and $d\in(-\infty,d^-)\cup(d^+,0)$ respectively (with
$-\infty < d^*, d^-, d^+<0$). In all cases we find convex stretches of the front with well defined
tangents limited by points with sharp edges. Using $\lambda = -d/(1-d)$ we reveal the intervals
$\lambda \in (\lambda^*, 1)$, $\lambda \in (0, \lambda^*)$, and $\lambda \in (0,
\lambda^+)\cup(\lambda^-, 1)$ respectively. For these intervals a series of SOO problems exist whose
solutions are always found where the $\tau_{\lambda} (\tilde{\Omega})$ match the tangent of the
front. These can be smoothly visited as $\lambda$ changes infinitesimally slow, just as before.

      Consider now figure \ref{fig:3}{\bf a} for $\lambda \in (0, \lambda^*]$. We can define SOO
problems in this range, but an abrupt ending of the front (nowhere does its slope match
$d=-\lambda/(1-\lambda)$ for $\lambda \in (0, \lambda^*]$) implies that the solution to all these
SOOs is the same. This is indicated by the gray fan in figure \ref{fig:3}{\bf a}: a collection of
isoenergetic surfaces ($\tau_\lambda(\Omega)$) with different inclinations has been pushed all the
way against the front arriving to one same solution. This happens also in figures \ref{fig:3}{\bf
b-c}: several isoenergetic surfaces (those with $\lambda \in [\lambda^*, 1)$ and $\lambda \in
[\lambda^+, \lambda^-]$ respectively) reach the same solution when pushed against the front. As we
vary $\lambda$ within these relevant intervals any order parameter remains unchanged (${d\theta\over
d\lambda}=0$) because we report persistently the same solution. But these same order parameters
change at non-zero rates as we approach $\lambda^{*,\pm}$ from the outside. Because every point of
the front can be reached for some $\lambda$, plotting $\theta$ (figure \ref{fig:3}, insets) produces
a continuous curve while there must be a discontinuity in the derivatives. This is the fingerprint
of {\bf second order phase transitions} and is coherent with an interpretation based on the
thermodynamic Gibbs surface \cite{Gibbs1873b, Maxwell1904}.

    \paragraph{First order phase transition}

      In thermodynamics, discontinuous transitions and metastability are associated to concavities
in the Gibbs surface \cite{Gibbs1873b, Maxwell1904}. The same is true about Pareto optimal designs,
for which the Pareto front is equivalent to the Gibbs surface.

\begin{figure*}[htbp]
          \begin{center}
            \includegraphics[width= 15 cm]{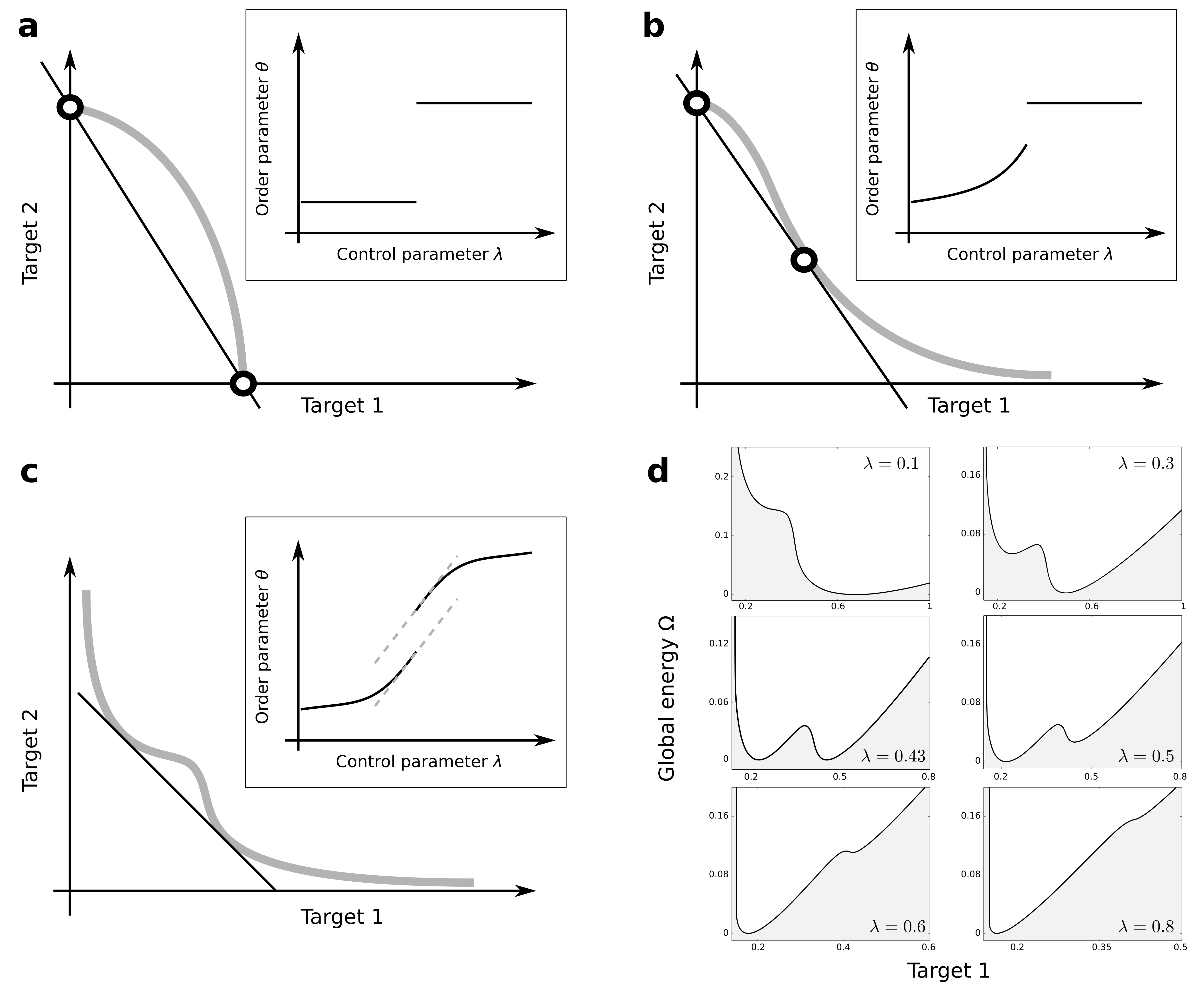}

            \caption{\textbf{Concave Pareto front, or fronts with concavities.} {\bf a} Only two
solutions are ever SOO global optima in a concave front: one if $\lambda>\lambda^*$ and another if
$\lambda<\lambda^*$. For $\lambda=\lambda^*$ both solutions coexist. Any order parameter presents a
sharp discontinuity at $\lambda=\lambda^*$. A similar situation happens in {\bf b} and {\bf c}. In
the later case, while $\theta(\lambda)$ is not continuous, its derivative is. {\bf d} An {\em
energetic landscape potential} is built through equation \ref{eq:2.01.01}. Plotting this function
for all the points of the front in {\bf c} reveals an energetic boundary below which no solutions
exist. Pareto suboptimal solutions lay above the boundary and SOO optima at fixed $\lambda$ sit at
the bottom of energy wells. Metastable solutions are associated to local minimums and lead to
hysteresis if we change $\lambda$ back and forth. }

            \label{fig:4}
          \end{center}
        \end{figure*}

      In a fully concave front (figure \ref{fig:4}{\bf a}), the straight line that joins both ends
of the front has slope $d^*$ and defines a critical value $\lambda^*\equiv{-d^*\over 1 - d^*}$. The
solution to any SOO with $\lambda < \lambda^*$ sits at the bottom-right end of the front. For
$\lambda > \lambda^*$ the solution lays at the top-left end. Both extremes coexist for $\lambda =
\lambda^*$. We cannot roll smoothly over such front by varying $\lambda$. A sudden shift between
radically different optima happens at $\lambda^*$. Any order parameter remains constant below and
above $\lambda^*$ but a gap exists between both constant values (figure \ref{fig:4}{\bf a}, inset),
as in first order phase transitions. Similar gaps are revealed when concavities are embedded within
convex stretches of the front (figure \ref{fig:4}{\bf b-c}). Global optima always lay on the convex
hull of the Pareto front. Pareto optimal designs inside the cavity might be metastable -- i.e. local
optima -- but are never global SOO optima. Metastability leads to the existence of hysteresis loops.

      A final illustration of first order phase transitions comes through the energetic landscape
enforced by $\Omega(x, \lambda)$ (figure \ref{fig:4}{\bf d}). For different values of $\lambda$ we
computed $\Omega(x, \lambda)$ for every Pareto optimal solution of the front in figure
\ref{fig:4}{\bf c}. This renders an energetic boundary (thick black curves) below which feasible
solutions do not exit (gray areas cannot be accessed). Heavy marbles rolling down the potential
wells minimize their energy. Metastability and hysteresis dynamics are due to changes in the
potential landscape (i.e. in the underlying SOO problem): As $\lambda$ varies a new pocket becomes
locally stable (figure \ref{fig:4}{\bf d}, $\lambda = 0.3$) and grows until it becomes the global
minimum ($\lambda = 0.43$ through $\lambda = 0.5$). We might get stuck in the new local minimum
until it is destabilized ($\lambda = 0.8$).

    \paragraph{Other situations}

      Phase transitions are among the most interesting phenomena accounted for by the Pareto front,
which generalizes the Gibbs surface for the studied MOO-SOO systems. Because of the equivalence
between both objects, we remit the reader to more specialized literature to discuss the more
technical details \cite{Varchenko1990, Aicardi2001, BouchetBarre2008}.

\begin{figure*}[htbp]
          \begin{center}
            \includegraphics[width= 15 cm]{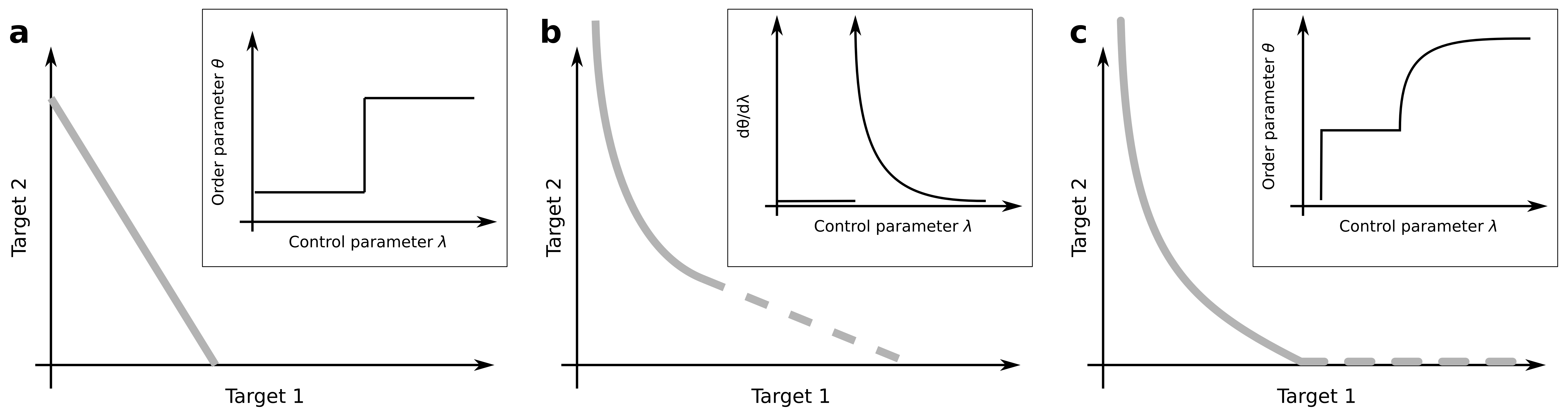}

            \caption{\textbf{A straight segment in the front resembles criticality.} {\bf a} A
situation similar to first order transitions for $\lambda \lessgtr \lambda^c$. At $\lambda^c$ the
whole front is SOO optimal. {\bf b} Degenerated SOO solutions can also happen at second order phase
transitions implying a diverging derivative in any order parameter. {\bf c} The dashed segment is
not Pareto optimal but it becomes SOO optimal if $\lambda = 0$ is allowed. }

            \label{fig:5}
          \end{center}
        \end{figure*}

      If the front contains a straight segment in the feature plane we define a critical value
$\lambda^c\equiv{-d^c\over 1 - d^c}$ with $d^c$ the slope of the segment. In figure \ref{fig:5}{\bf
a} a situation similar to the first order phase transitions from figure \ref{fig:4}{\bf a} is found
with either extreme of the front solving the SOOs defined by $\lambda \gtrless \lambda^c$
respectively. Besides, any Pareto optimal design is SOO optimal at this critical value. Hence, at
$\lambda = \lambda^c$ a plethora of very heterogeneous solutions not visited under any other
circumstance becomes available. Straight stretches of the front might also happen along second order
phase transitions (figure \ref{fig:5}{\bf b}). In all these cases the derivative of order parameters
with respect to $\lambda$ diverges for $\lambda \rightarrow \lambda^c$, as in critical points.

      $\lambda=0$ is the only situation in which solutions not in the Pareto front solve an SOO
(figure \ref{fig:5}{\bf c}). At least one of these solutions must be Pareto optimal. This case
is straightaway incorporated in the general framework, so allowing $\lambda_k=0$ does not alter our
theory.

      We assumed that the $\lambda_k$ are the only relevant control parameters, but other external
variables may modify the shape of the front or its elements. This could prompt phase transitions
into existence or erase them. (This can also be due to the varying of some $\lambda_k$ if $K>2$, as
noted for the Gibbs surface \cite{BouchetBarre2008}.) Parameters changing the constituents of the
front could trigger drastic changes not studied here.

    \subsection{Thermodynamics as a multiobjective optimization problem}
      \label{sec:3.02}

      We intend that phase transitions for MOO-SOO problems are as firmly grounded as those in
statistical mechanics, so we proceed now to show how thermodynamics is included in our theoretical
framework. Take an arbitrary physical system that can occupy any state $\sigma_j$ of an arbitrary,
abstract space $\sigma_j \in \Sigma$. Each $\sigma_j$ is a physical configuration with energy $E_j$.
Consider an arbitrary ensemble for this system $P_i$, in which $\sigma_j$ shows up with probability
$P_i(\sigma_j)$. Consider, indeed, all possible ensembles $P$ ($P_i\in P$), each of them an
arbitrary, mathematically consistent probability distribution, i.e. $$\sum_j P_i(\sigma_j)=1,$$ over
the space $\Sigma$. We define the functions:
        \begin{eqnarray}
          U(P_i, \Sigma) &=& \sum_j P_i(\sigma_j)E_j, \nonumber \\
          S(P_i, \Sigma) &=& - \sum_j P_i(\sigma_j) log(P_i(\sigma_j));
          \label{eq:2.02.01}
        \end{eqnarray} 
i.e. the internal energy and entropy of each ensemble $P_i$. These functions are rigorously defined
irrespective of whether they bear any physical meaning. Since the $P_i$ are arbitrary probability
distributions there is not any guarantee (neither necessity, so far) that $U(P_i, \Sigma)$ or
$S(P_i, \Sigma)$ obey any relevant relationship.

      These functions map $P_i \in P$ into the $U-S$ plane ($\mathbb{R}^2$), where dominance and
Pareto optimality are well defined. We can find the subset $\Pi \subset P$ of probability
distributions $P_\pi \in \Pi$ that minimize $U(P_i, \Sigma)$ and maximize $S(P_i, \Sigma)$
simultaneously. This is a legitimate MOO problem, again irrespective of whether it has got any
physical relevance. The only difference with earlier MOOs is that one of the targets is maximized,
which does not alter any of our conclusions. The solution to this problem ($P_\pi \in \Pi$)
constitutes the optimal tradeoff between the targets in equation \ref{eq:2.02.01}. This reduces the
number of relevant ensembles for us, but still there is no guarantee nor any need that these $P_\pi
\in \Pi$ present notable physical properties. They are just probability distributions solving an ad-
hoc MOO.

      Consider now the family of SOOs defined by:
        \begin{eqnarray}
          &\min \Big\{ \Omega(U, S; \lambda_U, \lambda_S) \equiv \lambda_U U + \lambda_S S \Big\} \Rightarrow \nonumber \\
          &\Rightarrow min\left\{ \hat{\Omega} \equiv {\Omega \over \lambda_U} = U + {\lambda_S \over \lambda_U} S \right\}. 
          \label{eq:2.02.02}
        \end{eqnarray}
This collapses the original MOO into a series of SOOs whose solutions lay upon the convex hull of
$\Pi$ in the $U-S$ plane, as shown above, so that phase transitions arise for singular values
$(\lambda_U/\lambda_S)^{*,\pm}$ due to concavities and sharp edges of the corresponding front. These
are still phase transitions of a fabricated problem.

      We leave this artificial MOO aside for a while. We can find the ensembles that maximize $S$
for fixed values of $U$, as dictated by the second law for equilibrium thermodynamics. These
distributions correspond to the microcanonical ensembles that can be mapped into the $U-S$ plane
through equation \ref{eq:2.02.01} describing a curve. For each fixed $U$ we attain the maximum
possible $S$, thus reconstructing the Pareto front of the MOO above (see SM). Irrespective of
whether thermodynamics consists in an MOO, microcanonical ensembles are linked to the Pareto front
of a legitimate MOO.

      The laws of thermodynamics also imply that 
        \begin{eqnarray}
          F = U - TS = U - {S/\beta} 
          \label{eq:2.02.03}
        \end{eqnarray}
is minimized in equilibrium at fixed temperature \cite{Maxwell1904}. Thermodynamic canonical
ensembles embody this minimization. Identifying $\hat{\Omega} \equiv F$ and $\lambda_S / \lambda_U
\equiv -T = -1/\beta$ from equations \ref{eq:2.02.02} and \ref{eq:2.02.03}, the relevant
$(\lambda_U/\lambda_S)^{*,\pm}$ correspond to those temperature values at which phase transitions
occur. Irrespective of whether thermodynamics consists of an MOO-SOO collapse, canonical ensembles
are constrained by the rules that reveal phase transitions in one such system. First order phase
transitions are associated to cavities at which the Pareto front and its convex hull (i.e.
microcanonical and canonical ensembles) must differ. This is in agreement with recent literature in
{\em ensemble inequivalence} \cite{TouchetteTurkington2004} some of whose results can be intuitively
derived in our framework.

      As a final remark, thermodynamic systems described by internal energy, entropy, and volume
find their microcanonical ensembles laying at the more general surface defined by the Gibbs
potential $G = U - TS - pV$ \cite{Gibbs1873b, Maxwell1904}. This also corresponds to the Pareto
front of an MOO problem. Phase transitions are then identified for singular values of temperature
and pressure (see SM).

\begin{figure*}[htbp]
          \begin{center}
          \includegraphics[width= 15 cm]{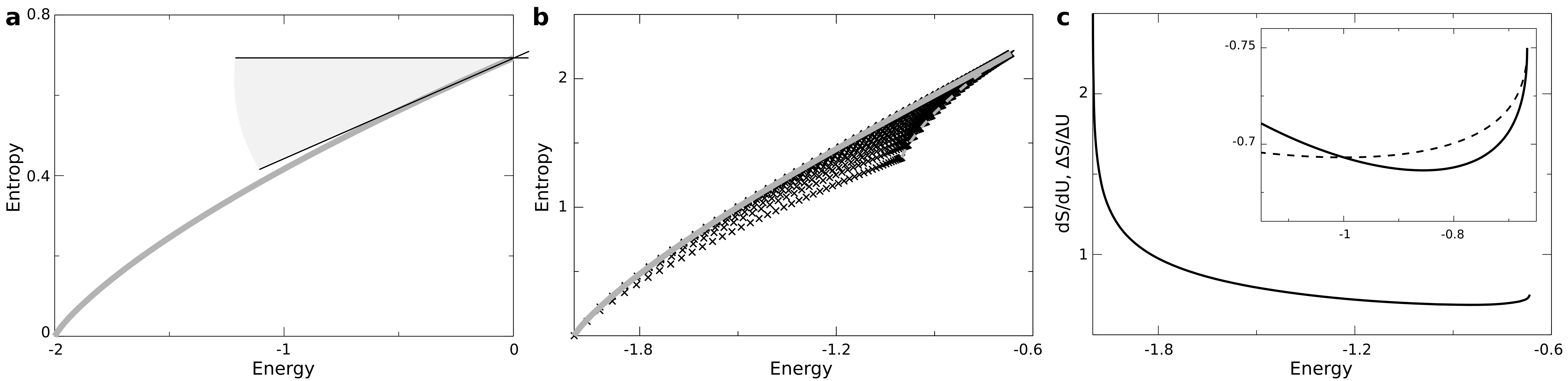}

            \caption{\textbf{Pareto fronts for the Ising and Potts models.} {\bf a} The front of the
mean-field Ising model (thick gray line) is convex but ends abruptly revealing a range $\beta \in
(0, 1/Jz]$ (gray fan) for which SOOs arrive to the most entropic solution always. {\bf b} A sample
of arbitrary distributions $P=\{p_1,p_2,p_3\}$ (black crosses) for the $q=3$ Potts model is
dominated by its Pareto front whose top-right part is concave. This indicates a first order phase
transition. {\bf c} That cavity becomes noticeable when analyzing the slope of the front, which is
not monotonously decreasing. }

            \label{fig:6}
          \end{center}
        \end{figure*}

      The Ising and Potts models illustrate second and first order transitions respectively. General
versions of these models have been solved using ensemble inequivalence \cite{BertalanNishimori2011,
CosteniucTouchette2005}. They are discussed here because they allow an almost complete analytic
resolution using the methodology and vocabulary of MOO (see SM for details). The Ising model
presents a convex Pareto front (figure \ref{fig:6}{\bf a}). This front results in a function
$S=\bar{S}(U)$ well defined for $U\in[-Jz/2, 0]$, $J$ being the coupling constant of the model and
$z$ the number of nearest-neighboring spins. Because entropy is maximized (as opposed to the
minimizations in previous sections) the Pareto front has positive slope. Apart from this, exactly as
solutions for SOOs with fixed $\lambda$ were found where the slope $d$ of the front matched $d\equiv
-\lambda/(1-\lambda)$, now solutions for fixed $\beta$ are associated to some slope $d(\beta)$. We
get curves of constant free energy $\tau_\beta(F)$ just as we got isoenergetic surfaces
$\tau_\lambda (\Omega)$ before:
        \begin{eqnarray}
          \tau_\beta(F) &=& \Big\{ (U, S) | S = \beta U - \beta F \Big\}. 
          \label{eq:2.02.04}
        \end{eqnarray}
The free energy minimization is equivalent to pushing these $\tau_\beta(F)$ as much as possible
against the front without changing its slope (i.e. without changing $\beta$, thus at constant
temperature). This reveals how SOO solutions are found now precisely where ${\partial S/\partial U}
= d = \beta$. The derivative $\partial \bar{S}/\partial U$ tends to $+\infty$ for $U\rightarrow
(-Jz/2)^+$, meaning that this end of the front is only reached for $\beta \rightarrow +\infty$ (i.e.
$T \rightarrow 0$). Nothing remarkable happens there. At the other end of the front $\partial
\bar{S}/\partial U$ tends to $1/Jz$ for $U \rightarrow 0^-$. Thus, the range $\beta \in (1/Jz, 0]$
(gray fan in figure \ref{fig:6}{\bf a}) defines SOOs whose solutions are always the most entropic
configuration of the Ising model. At $\beta\ge1/Jz$ the SOO solution leaves the disordered phase
towards the ordered, magnetic one.

      The Potts model (its front and a sample of not Pareto-optimal solutions is shown in figure
\ref{fig:6}{\bf b}) presents a concavity in the top-right end of its Pareto front as revealed
by the derivative $dS/dU$, which is not monotonously decreasing (figure \ref{fig:6}{\bf c}).
This implies a first order phase transition from the most entropic configuration to the ordered
state, as it is known \cite{KiharaShizume1954}. Calculations based on the convex hull of this front
match those obtained elsewhere \cite{Wu1982} (see SM). Again, nothing remarkable happens at the
lower left end of the front, which is reached for $\beta \rightarrow +\infty$ ($T \rightarrow
0$).

  \section{Discussion}
    \label{sec:04}

    In this paper we presented a theoretical framework to discuss Pareto optimal systems that
collapse into families of SOOs. We have uncovered deep equivalences of key objects in MOO (the
Pareto front) and fundamental concepts of thermodynamics (the Gibbs surface) \cite{Gibbs1873a,
Gibbs1873b, Maxwell1904}. Consequently, a firm basis exists to extend this physical theory to a
wider collection of MOO-SOO systems.

    An important motivation of our paper was to find broad regularities based on the general
structure of the optimization problems, without regard of what is being optimized in each particular
case. Given two different problems that happen to have the same Pareto front, we automatically know
that both systems undergo the same phase transitions: in both systems they will have the same order
(first or second) and location in control parameter space. Only their physical or geometrical
interpretations will differ. The most important contribution of this paper is from thermodynamics
towards the investigation of Pareto optimal systems. The possibility of defining effective
potentials associated to the Pareto front go hand in hand with the first and second order
transitions revealed.

    By considering the theoretical framework developed in this paper, these kind of analyses are
easily available (and absent, to the best of our knowledge) for other Pareto optimality research.
This, we believe, enriches the interpretation of Pareto efficient systems. Take the example
mentioned above involving the minimization of short- vs. long-range forces in protein structure
\cite{CutelloNicosia2006}. The fronts shown in figure \ref{fig:1}{\bf d-e} strongly suggest the
existence of first order phase transitions between different protein configurations. These
transitions are intrinsically physical but can be located thanks to the Pareto formalism. Other
authors have studied the tradeoffs related to information spread in complex networks
\cite{GoniSporns2013}. Different network topologies minimize/maximize the efficiency of routing or
enhance/hinder the diffusion of information. Phase transitions are also evident in this example (see
figure 3 in \cite{GoniSporns2013}). We can consider the entropy associated to the information routed
through the network. That entropy can eventually be linked to energy through the Landauer principle
\cite{Landauer1961} so that these transitions might have clear physical interpretations. 

    A review of MOO literature from the proposed perspective seems necessary now, but each
individual case should be studied carefully. Very recent contributions are exploring biology from an
MOO perspective, making it a very attractive territory where to apply our insights. But this field
can also prove challenging. Following \cite{ShovalAlon2012} and \cite{SzekelyAlon2013}, several
tasks ($P_1, ..., P_K$) can contribute to the fitness of a species through $F=F(P_1, ..., P_K)$.
Improvements in any $P_k$ might raise the overall fitness justifying an MOO study. But the
consequences of the MOO-SOO combination have not been explored. Our framework relies on the
integration of the targets that in the case of thermodynamics and other problems happens through a
linear combination (as in equation \ref{eq:2.01.01}). The scenario suggested in
\cite{ShovalAlon2012, SzekelyAlon2013} becomes interesting because the MOO-SOO collapse imposed by
$F=F(P_1, ..., P_K)$ is not necessarily linear. In our framework SOO optima are always mutually 
non-dominating so that SOO and MOO problems can be simultaneously solved. This might not be the case
for non-linear fitness functions, opening interesting possibilities associated to biological
fitness. Further research should address these and other pending issues.

  \section*{Acknowledgments}

    We thank the members of the CSL for useful discussions. This work was supported by grants from
the Fundaci\'on Bot\'in, the European Research Council (ERC Advanced Grant) and by the Santa Fe
Institute.

\appendix

  \section{A discussion of thermodynamics as an MOO problem}
    \label{sec:SM1}

    We have shown how the solutions to an MOO-SOO problem based on a physical system correspond
precisely to the canonical ensembles of that system in thermodynamic equilibrium. A crucial step is
proving that the microcanonical ensemble reconstructs the Pareto front of the relevant MOO. We
devote some space here to better clarify this point.

\begin{figure*}[htbp]
        \begin{center}
          \includegraphics[width= 15 cm]{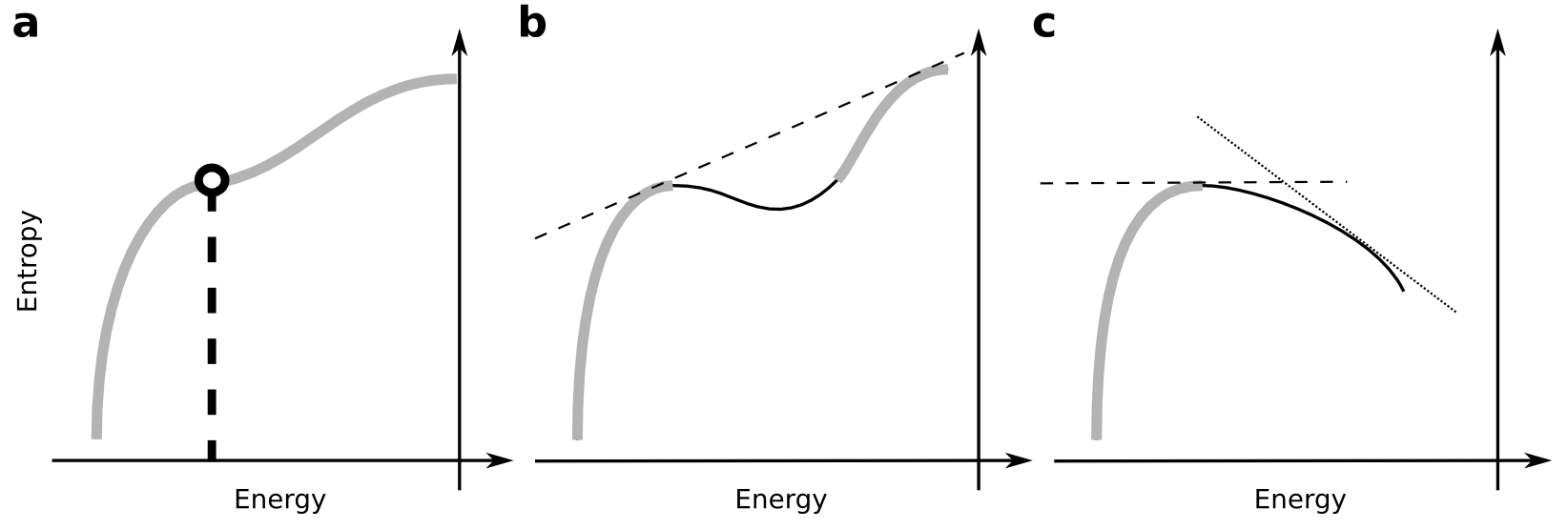}

          \caption{\textbf{Laws of thermodynamics and the Pareto front.} {\bf a} According to the
second law of thermodynamics, at constant internal energy (vertical dashed line) the microcanonical
ensemble is the one that maximizes the entropy (and is hence mapped into the open circle in this
example). Implementing this maximization for varying energy yields a function on the $U-S$ space
that for thermodynamic systems is usually monotonously increasing with $U$ -- more energetic systems
usually have more entropy. This guarantees that any two points on this curve are mutually  non-
dominated. There cannot be any point above this curve, thus the obtained curve must be exactly the
Pareto front of the corresponding MOO problem. {\bf b} This curve would not match the front only if
the microcanonical entropy were not monotonously increasing with $U$. This is an odd situation in
thermodynamics. These non-increasing stretches would necessarily lay inside a cavity (solid black
curve) and would never show up in thermodynamic equilibrium. {\bf c} Such situation can also happen
beyond the global maximum of the entropy, which is only reached for $T=0$ (dashed line). Points of
the microcanonical entropy beyond this maximum would require $\partial S / \partial U = 1/T<0$ (i.e.
negative temperatures, dotted line). In both {\bf b} and {\bf c} the entropy of microcanonical
ensembles still contains the whole Pareto front and, of course, its convex hull. }

          \label{fig:SM1}
        \end{center}
      \end{figure*}

    The reconstruction of the Pareto front through microcanonical ensembles derives from the second
law of thermodynamics. Microcanonical ensembles are those that maximize the entropy for a fixed
value of internal energy (figure \ref{fig:SM1}{\bf a}). Each microcanonical ensemble is mapped into
the $U-S$ plane through equations \ref{eq:2.02.01}. All these ensembles together trace a curve that
must be a function $S = \bar{S}(U)$. Otherwise, for a value of $U$ two or more values of $S$ would
be assigned, and only one of them could me maximum. The lower values would not correspond to any
microcanonical ensemble, accordingly.

    Consider the situation in which $S=\bar{S}(U)$ increases monotonically. Then a greater internal
energy is always associated to a greater entropy, which intuitively makes sense. This bijectivity
guarantees that $S=\bar{S}(U)$ matches the corresponding Pareto front. For each $U_j > U_i$
necessarily $\bar{S}(U_j) > \bar{S}(U_i)$, thus in this curve there are not any two $U_i < U_j$ such
that $\bar{S}(U_i) > \bar{S}(U_j)$, which would imply $U_i \prec U_j$. Furthermore, any ensemble
mapped into a point $(U_i, S_i)$ outside this curve is necessarily dominated by some microcanonical
ensemble mapped into $(U_i, \bar{S}(U_i))$ since by definition microcanonical ensembles are such
that $\bar{S}(U_i)>S_i$ for that given value $U_i$. Summing up: i) points along the curve
$S=\bar{S}(U)$ are mutually non-dominated and ii) for any physically plausible point $(U, S)$
outside this curve there is at least one point that belongs to the curve and dominates $(U, S)$.
This is the definition of the Pareto front indeed. Once microcanonical ensembles reconstruct the
front of the proposed MOO, the collapse into SOOs and phase transitions due to the shape of the
front follow naturally because the free energy is minimized in thermodynamic equilibrium.

    Some mathematical idealizations of physical systems admit non-monotonically increasing
entropies. These often lead to exotic parameterizations such as negative temperatures. Furthermore
they do not affect the current theory as we will show now. Consider figures \ref{fig:SM1}{\bf b} and
{\bf c} regardless of the physical reality of such descriptions. The equilibrium thermodynamics of
such hypothetical systems are still well represented by the convex hull of the Pareto front, hence
out theoretical framework remains true. Given the definition of dominance, we note that the Pareto
front of the relevant MOO problem is still fully reconstructed by the curve $S=\bar{S}(U)$ (thick
stretches in figures \ref{fig:SM1}{\bf b-c}). Non-increasing stretches of $S=\bar{S}(U)$ lay inside
a cavity (figure \ref{fig:SM1}{\bf b}) or after the global maximum of the function (figure
\ref{fig:SM1}{\bf c}). If they are inside a cavity, such situations never show up in thermodynamic
equilibrium, whose canonical ensembles are strictly mapped into the convex hull of the front. These
points are bypassed by a first order phase transitions. In the other situation, the slope of the
Pareto front at the global maximum (whose limit from below is perfectly reconstructed by the
microcanonical ensemble) is necessarily $0$, meaning that such situation is only reached at $\beta
\rightarrow 0 \Rightarrow T \rightarrow +\infty$. Reaching solutions beyond the global maximum
implies considering $\beta = 1/T < 0$, which is not realistic. Thus non monotonously increasing
functions $S=\bar{S}(U)$ do not affect the general framework because they are situations that we are
not concerned with in thermodynamic equilibrium. For MOO-SOO systems others than thermodynamics we
do not rely on microcanonical ensembles, but on the Pareto front straightaway, in which 
non-dominated regions (including curves that would break the monotonic trend of the front) never
show up.\\

    Our work follows closely some ideas from Gibbs \cite{Gibbs1873a, Gibbs1873b, Maxwell1904} that
did not become so mainstream in the study of statistical mechanics. Gibbs's {\em graphic method}
relied on the existence of a surface upon which all possible states of a thermodynamic species in
equilibrium dwell. This surface is defined by the Gibbs potential $G(p,T) = U + pV - TS$, which
plays the role of equation \ref{eq:2.01.01} both in thermodynamics and in our theoretical framework.
We identify the target functions $U$, $S$, and $V$ and the control parameters $T$ and $p$. The
tangent plane at a given point of the surface is defined by a normal vector whose components are
precisely related to the pressure and temperature of that equilibrium state \cite{Maxwell1904}.

    It was argued that an MOO approach might be adequate when the different targets cannot be
compared (e.g. if they have different units). In thermodynamics, the control parameters $T$ and $p$
transform different potentials into the same units. At fixed temperature entropy is heat and at
fixed pressure volume is work -- hence a low entropy and an unoccupied volume are available free
energy. All free energy must be utilized to reach the thermodynamic equilibrium \cite{Maxwell1904}
so that $G(p, T)$ is minimized and only the convex hull of the Gibbs surface shows up. When changing
$T$ or $p$ quasistatically, concavities are bypassed revealing first order phase transitions. Sharp
edges are consistently related to second order phase transitions again. 

    This ingenious picture received renewed attention recently \cite{Varchenko1990, Aicardi2001} and
is connected to the concept of ensemble inequivalence \cite{TouchetteTurkington2004}. The Gibbs
surface is fabricated thanks to the microcanonical ensemble and it can be convex or concave, while a
thermodynamic canonical ensemble can only be convex (note that we use a different convention for
concavity/convexity). Whenever $G$ becomes concave both ensembles must diverge geometrically in the
$U-V-S$ space. This makes the canonical ensemble non-analytic at the inequivalence points which is
reflected as a first order phase transition in the corresponding physical system.

  \section{Solving the Ising and Potts models from an MOO perspective}
    \label{sec:SM2}

    As noted in the main text, the Ising and the Potts models and more general versions of them have
been solved using the concept of ensemble inequivalence \cite{BertalanNishimori2011,
CosteniucTouchette2005}, which is comprehensively explained by the Pareto optimality approach
introduced here. We illustrate thermodynamic phase transitions with the Pareto front through these
models because of their historical importance and because they allow a complete analytical
treatment. The details of the calculations follow. 

    \paragraph{The mean-field Ising model -- a second order phase transition}
      \label{app:2.01}

      We use a standard mean-field Hamiltonian for the Ising model $H_j = - {J \over 2}
\sum_{\left<j,k\right>} s_js_k$ with $J$ the coupling constant and the sum running over $z$
neighboring spins. We parameterize the system with the probability $p$ that we find the mean-field
spin in the {\em up} state. It becomes easy to write down the entropy and the internal energy of the
system in terms of $p$:
        \begin{eqnarray}
          S &=& -p \; log(p) - (1-p) \; log(1-p), \nonumber \\ 
          U &=& -{Jz\over 2}(2p-1)^2. 
          \label{eq:app2.01.01}
        \end{eqnarray}
Solving this last expression for $p$, we can also write the entropy as a function of the internal
energy alone:
        \begin{eqnarray}
          \bar{S}(U) &=& -{1\over 2}\left\{ 1+f(U) \right\} log\left( {1+f(U)\over 1-f(U)} \right) \nonumber \\
          && - log\left( {1-f(U)\over 2} \right), 
          \label{eq:app2.01.02}
        \end{eqnarray}
with: 
        \begin{eqnarray}
          f(U) &\equiv& +\sqrt{{-2U\over Jz}}. 
          \label{eq:app2.01.03}
        \end{eqnarray}

      Equation \ref{eq:app2.01.02} (represented in figure \ref{fig:6}{\bf a}) gives us $S$ as a
function of $U$ ($S=\bar{S}(U)$) for all possible states that the model can be found into, disregard
of whether or not these states correspond to thermodynamic equilibrium situations or to
microcanonical ensembles. Because for this model equation \ref{eq:app2.01.02} is a function, for
each $U$ $\bar{S}(U)$ is also maximal -- i.e. every state of the system corresponds to a
microcanonical ensemble itself. This is luckily valid for this one model, but not necessarily true
in a general case.

      This curve also constitutes the Pareto front of the corresponding MOO problem (minimizing $U$
and maximizing $S$ from equations \ref{eq:2.02.01}), as is expected given the correspondence between
the microcanonical ensembles and Pareto optimal solutions. Let us analyze the Pareto optimality of
equation \ref{eq:app2.01.02}. First we note that $\bar{S}(U)$ is only real and well defined for
$U\in[-Jz/2,0]$, the range of available energies for the model. In this range:
        \begin{eqnarray}
          {d\bar{S}\over dU} &=& -{1\over 2}f'(U)log\left( 1+f(U)\over 1-f(U) \right) > 0, 
          \label{eq:app2.01.04}
        \end{eqnarray}
thus $\bar{S}(U)$ is monotonously increasing, which guarantees that its points in the $U-S$ plane do
not dominate each other regarding energy minimization and entropy maximization. Because this curve
comprises everything that is possible in the system under research and because its constituting
points are mutually non-dominated, it must be the Pareto front itself.

      Besides, ${d\bar{S}\over dU}$ is positive and monotonously decreasing for $U\in(-Jz/2,0)$;
thus there are not concavities in the Pareto front: we rule out first order phase transitions. We
can also rule out second order phase transitions in the interior of $U\in(-Jz/2,0)$ because the
derivative is well defined everywhere. Second order phase transitions are thus restricted to
$U=-Jz/2$ or $U=0$. We inspect ${d\bar{S}\over dU}$ as $U$ tends to these points. After some algebra
we arrive to:
        \begin{eqnarray}
          \lim_{U\to 0^-} {d\bar{S}\over dU} &=&  {1\over Jz}. 
          \label{eq:app2.01.05}
        \end{eqnarray}
and: 
        \begin{eqnarray}
          \lim_{U\to (-Jz/2)^+} {d\bar{S}\over dU} &=& +\infty. 
          \label{eq:app2.01.06}
        \end{eqnarray}

      As we saw in the results section, for the SOO posed at temperature $T=1/\beta$ we typically
reach points of the Pareto front where their tangent matches the slope of the corresponding
$\tau_\beta(F) = \left\{ (U, S) | S = \beta U - \beta F \right\}$. The derivative ${d\bar{S} \over
dU}$ as we approach $U=-Jz/2$ is infinite, meaning that we will reach this end of the Pareto front
only at $\beta\rightarrow \infty$ (zero temperature). There is not any remarkable behavior here. At
the other end of the Pareto front the derivative is not $0$, but a finite positive number. This
means that already for $\beta={1\over Jz}$ the free energy optimum -- i.e. the SOO solution -- is
located at the upper right end of the front which corresponds to the state with more entropy and
energy. If we further decrease $\beta$ we will not be reaching any novel solutions: the SOO optima
remain the most entropic state of the system. Going back to the well behaved range of $\beta$, as we
increase it above ${1\over Jz}$ the SOO solutions begin to roll over the Pareto front continuously.
The transition between a persistent solution for $\beta\in (0,1/Jz]$ and a varying solution in the
regime $\beta\in(1/Jz, +\infty)$ implies a discontinuity in the derivative of any order parameter.
This is analogous to the cases illustrated in figure \ref{fig:3} of the main text, and is associated
to second order phase transitions as the one that we know that the mean-field Ising model undergoes
at precisely $\beta=\beta^c \equiv {1\over Jz}$.

    \paragraph{The mean-field Potts model -- a first order phase transition. }
      \label{app:2.02}

      We repeat the same operations with the Bragg-Williams approximation to the Potts model. This
has been solved somewhere else \cite{KiharaShizume1954} using other methods. This choice of
implementing the mean-field presents first order phase transitions for any $q\ge3$, where $q$ is the
number of available states for each spin. For a discussion of the Braggs-Williams against other
mean-field approaches to the Potts model see \cite{Wu1982}.

      Following \cite{KiharaShizume1954}, we write down the entropy and energy of the system:
        \begin{eqnarray}
          S &=& -\sum_{j=1}^q p_j log(p_j), \nonumber \\ 
          U &=& -{zJ\over 2} \sum_{j=1}^q p_j^2;
          \label{eq:app2.02.01}
        \end{eqnarray} 
with $J$ and $z$ still the coupling and the number of neighbors. Now $U$ and $S$ are parameterized
by the probabilities $p_j$ ($j=1,...,q$) of finding a spin in each of the $q\ge3$ states. The
normalization $\sum_j p_j=1$ means that there are $q-1$ parameters and we cannot write
$S=\bar{S}(U)$ as before unless we make some assumption. Let us prefer one arbitrary state (say
$j=1$) over the others. Let us call $p$ to the probability of finding a spin in that preferred
state, and let us further assume that any other state is equally likely $p_{j'}=(1-p)/(q-1)$, now
with $j'=2,...,q$. This is analytically justified in the literature \cite{KiharaShizume1954} and
later by our arguments about Pareto dominance. We note that, unlike for the Ising model, states
compatible with the premises of the system will not usually be constrained to a curve because we
have too many degrees of freedom. In figure \ref{fig:6}{\bf b} (main text) we represent a sample of
valid points for $q=3$: all of them can happen in theory (they are mathematically valid descriptions
of the system). A few of them Pareto dominate some others thus not all of these configurations will
be reached in thermodynamic equilibrium.

      Thanks to the previous symmetry breaking to favor one state over the others we can write down
the following curve:
        \begin{eqnarray}
          \bar{S}(U) &=& -{1+f_q(U)\over q} log\left( {(q-1)(1+f_q(U)) \over q-1-f_q(U)} \right) \nonumber \\ 
          &&- log\left( q-1-f_q(U)\over q(q-1) \right). 
          \label{eq:app2.02.02}
        \end{eqnarray}
This is the counterpart of equation \ref{eq:app2.01.02} only now: 
        \begin{eqnarray}
          f_q(U) &\equiv& \sqrt{(1-q)\left[ 1+{2qU \over zJ} \right]}. 
          \label{eq:app2.02.03}
        \end{eqnarray}
Equation \ref{eq:app2.02.02} is represented in figure \ref{fig:6}{\bf b} of the main text for $q=3$.
We can appreciate that it is monotonously growing as a function of $U$: its points are mutually
non-dominated and constitute the Pareto front. The fact that there is not any point in the previous
sample that Pareto dominates any point in this curve suggests that our symmetry breaking hypothesis
(favoring one spin state over the others) is correct. It can be analytically proved, indeed
\cite{KiharaShizume1954}. Although it is visually difficult to appreciate, a concavity exists in the
upper right part of the Pareto front. This becomes more obvious when analyzing ${d\bar{S}\over dU}$
(figure \ref{fig:6}{\bf c} of the main text), which is not monotonously decreasing.

      Most of the Pareto front is continually visited: as we vary $\beta$, the SOO solutions roll
over its convex lower-left part. It can be shown once more that the less energetic extreme of the
front is reached only for $\beta\rightarrow+\infty$ and $T=0$, so that there is not any remarkable
feature in that temperature range. Once again, it exists a value $\beta^*$ below which the solution
is persistently the most entropic one. At exactly $\beta = \beta^*$ that solution coexists with
another one in the convex part of $\bar{S}(U)$.

\begin{figure}[htbp]
          \begin{center}
            \includegraphics[width= 8 cm]{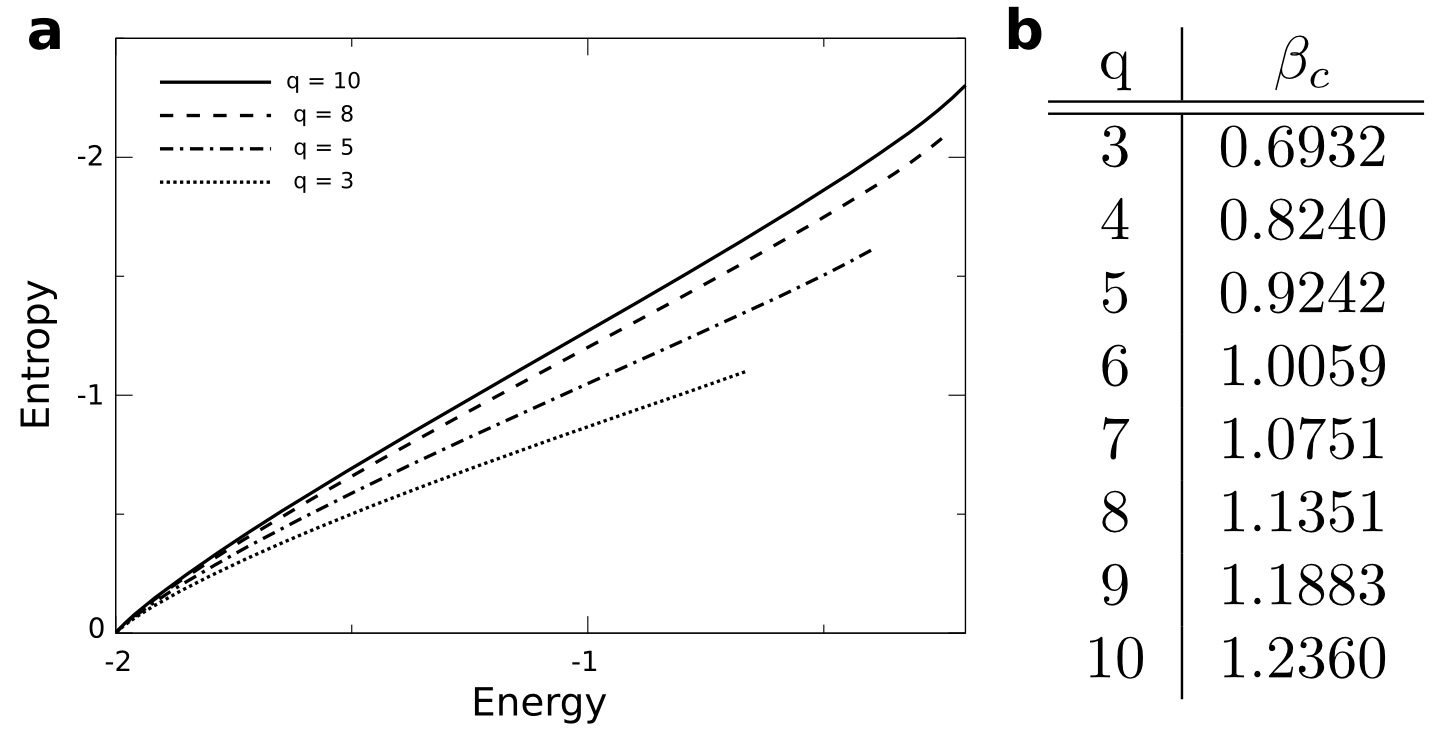}

            \caption{\textbf{Pareto front for the mean-field Potts model with different $q$.} {\bf
a} Pareto fronts of the Potts model for $q=3, 5, 8, 10$. Although hardly noticeable, all these
fronts have got concave stretches towards their upper-right ends. This indicates that all of them
undergo first order phase transitions from the most disordered state to a more ordered phase where
symmetry has been broken to favor just one of the states. A Pareto front for $q$ dominates the
Pareto front for every $q'<q$ indicating that this system will never leave empty one of its
available states spontaneously, except in the most ordered state in which all spins are aligned.
{\bf b} Inverse temperature at which the mean-field Potts model presents its first order phase
transition for $q=3,...,10$. The results match perfectly those from the literature
\cite{KiharaShizume1954}.}

          \label{fig:SM2}
          \end{center}
        \end{figure}

      To locate $\beta^*$ we plot ${d\bar{S}\over dU}$ and we compare it to the slope $\Delta S
\over \Delta U$ of the straight line that connects the top-right extreme with other points along the
Pareto front (main text, figure \ref{fig:6}{\bf c}, inset). Where both functions intersect we have
identified the phases that coexist. The straight line that connects these phases has slope $\beta^*$
precisely. We collect $\beta^*$ for the Potts models with different parameter values $q$ in figure
\ref{fig:SM2}{\bf b}. These results match those known from the literature \cite{KiharaShizume1954}.

      Because the two coexisting solutions are far away in the Pareto front, at $\beta=\beta^*$
these systems undergo a drastic change -- as opposed to the continuous transition from the Ising
model. This is similar to the first order phase transition situation illustrated in figure
\ref{fig:4} of the main text. This is the kind of phase transition that we know that our model
presents.

      In figure \ref{fig:SM2} we represent the Pareto front for many values of $q$. We
observe that solutions for $q$ dominate solutions for any $q'<q$, leading to the interesting (while
trivial) observation that an instance with higher $q$ does not spontaneously decay towards one with
less available states by setting some $p_{j'}=0$. This is well known for the Potts model, but such
an interesting scenario should not be discarded in general for a different problem, and the Pareto
front could provide a formalism to detect such a possibility.

\end{document}